\documentclass[graybox]{svmult}

\usepackage{mathptmx}       
\usepackage{helvet}         
\usepackage{courier}        
\usepackage{type1cm}        
\usepackage{makeidx}         
\usepackage{graphicx}        
\usepackage{multicol}        
\usepackage[bottom]{footmisc}

\makeindex             

\begin{document}

\title*{Dense Matter in Neutron Star: Lessons from GW170817}
\author{Sarmistha Banik and Debades Bandyopadhyay}
\authorrunning{Dense matter in neutron star} 
\institute{Sarmistha Banik, Birla Institute of Technology and Science Pilani, 
Hyderabad Campus, Hyderabad - 500078, India \email{sarmistha.banik@bits.hyderabad.ac.in} 
\and Debades Bandyopadhyay, Astroparticle Physics and Cosmology Division and
Centre for Astroparticle Physics, Saha Institute of Nuclear Physics, HBNI, 1/AF 
Bidhannagar, Kolkata-700064, India \email{debades.bandyopadhyay@saha.ac.in}}
%
%
\maketitle

\abstract{Neutron star merger event GW170817 sets an upper limit on the 
maximum mass of non-rotating neutron stars. Consequently, this event puts 
strong 
constraints on the dense matter equation of state (EoS). A comparative study
of dense matter equations of state (EoSs) is presented here. It is found that
the $\Lambda$ hyperon EoS BHB$\Lambda\phi$ (Banik, Hempel $\&$ Bandyopadhyay 
2014) constructed within the framework of the density dependent hadron field 
theory is favoured.}

\section{Introduction}
\label{sec:1}

S. Chandrasekhar predicted the mass limit for the first family of 
compact astrophysical objects known as White Dwarfs \cite{chandra31}. Next 
L.D. Landau should be credited for his idea about the second family of compact 
objects, as 'giant nucleus' \cite{land32}. After the discovery of neutrons, it 
was realised that the second family might be neutron stars \cite{bz34}. First 
pulsar was discovered in 1967 \cite{hewis68}. We are celebrating 50 years of 
the discover of first pulsar in 2017. What could be a better celebration than 
finding the neutron star merger event GW170817 \cite{abbott}. This stands out as
a very important discovery in the history of mankind.

Neutron star merger event GW170817 was discovered both in gravitational waves 
and light. The gravitational wave signal was observed in LIGO detectors
\cite{abbott}. 
A short Gamma Ray Burst (sGRB) was recorded 1.7 s after the merger 
by the Fermi-GBM \cite{abbott2}. This, for the first time, established a link 
between a neutron star merger event and sGRB. Later electromagnetic signals in 
visible, ultra-violate and infra-red bands were detected from the ejected matter
which formed a 'kilonova'. 

GW170817 is a boon to the nuclear astrophysics community because it 
allows to probe compositions and EoS in neutron star and r-process 
nucleosynthesis in the ejected neutron rich matter. 
The merger event provides crucial information about the remnant 
and neutron stars in the binary. The chirp mass is estimated to be
1.188$^{+0.004}_{-0.002}$ M$_{\odot}$. Assuming low spins as found from 
observations of neutron stars in our Galaxy, individual neutron star mass 
in the binary ranges 1.17-1.60 M$_{\odot}$. The massive remnant formed
in the merger has a mass 2.74$^{+0.04}_{-0.01}$ \cite{abbott}.  

The outstanding question is what happened to the massive remnant formed
in GW170817. The prompt collapse of it to a black hole is ruled out because 
large amount of matter was ejected. In this 
situation, either the remnant is a long lived massive neutron star or it 
collapsed to a black hole. Recent x-ray observation using the Chandra 
observatory indicates that the massive remnant might be a black hole 
\cite{pool}.  

It is possible to estimate the upper limit on the maximum maximum 
($M_{max}^{TOV}$) of
the non-rotating neutron star if the remnant becomes a black hole through 
delayed collapse. Different groups have determined the upper limit on 
$M_{max}^{TOV}$ from 
the multimessenger observation of GW170817 as well as from numerical 
relativity \cite{metz,rezo,shap}. All these estimates converge to the same 
value of $\sim 2.16$ M$_{\odot}$ for the upper limit on $M_{max}^{TOV}$. It is 
already known from the observations of neutron stars that the most massive
neutron star has a 2.01 M$_{\odot}$ which sets the lower limit on 
$M_{max}^{TOV}$ \cite{anto}. All these information tell us that the maximum 
mass of 
non-rotating neutron stars should be in the range $2.01<M_{max}^{TOV}<2.16$. 
This constraint on 
$M_{max}^{TOV}$ might severely restrict EoS models. This motivates us to carry 
out a comparative study of EoS models involving Banik, Hempel and Bandyopadhyay
(BHB) EoS
with hyperons in the density dependent relativistic hadron (DDRH) field 
theory \cite{typ,apj14}.   
  
%

We organise the article in the following way. We introduce the density dependent
hadron field theory and BHB$\Lambda \phi$ EoS in Section 2. Results are 
discussed in Section 3. We conclude in section 4.

\section{Equation of State for Neutron Star Matter}
\label{sec:2}
Equation of state is an important microphysical input for the study of 
core-collapse supernovae (CCSN), neutron stars and neutron star mergers
\cite{char15,rad17}. For
CCSN and neutron star merger simulations, an EoS is a function of three
parameters - density, temperature and proton fraction. These parameters vary 
over wide range of values. For example, density varies from $10^2 - 10^{15}$ 
g/cm$^3$, temperature from 0 to 150 MeV and proton fraction from 0 to 0.6. In
this study, we focus on neutron star EoSs which are derived from the EoS 
constructed for CCSN and neutron star merger simulations. Particular, we 
describe here the BHB EoS and adopt the same for our calculation \cite{apj14}.

The compositions of matter in CCSN and neutron star changes with density, 
temperature and proton fraction. Below the saturation density 
(2.7$\times 10^{14}$ g/cm$^3$) and low temperature, nuclei and nuclear clusters 
are present and make the matter inhomogeneous. In this case, non-uniform matter
is made of light and heavy nuclei, nucleons and leptons in thermodynamic 
equilibrium. Matter above the saturation density is uniform. Several novel 
phases of matter such as hyperons, kaon condensate or quarks might appear at 
higher densities. We discuss both (non-)uniform matter in the following 
subsections.        
\subsection{Non-uniform matter}
\label{subsec:2}
Here the in-homogeneous matter is described by an extended version of the 
Nuclear Statistical Equilibrium (NSE) model that was developed by Hempel and 
Schaffner (HS) \cite{hs}. The extended NSE model takes into account 
interactions among nucleons, interaction of nuclei or nuclear clusters with 
the surrounding medium. Furthermore, the Coulomb interaction is considered. 

Interactions among unbound nucleons are treated in the relativistic mean filed
(RMF) approximation using a density dependent relativistic hadron field 
theory. Nuclei are
considered as classical particles described by the Maxwell-Boltzmann statistics.
Binding energies of 
thousands of nuclei entering into the calculation are obtained from the nuclear 
mass data table \cite{audi03}. When experimental values are not 
available, theoretically calculated values are exploited \cite{moller95}. 
Medium modifications of nuclei 
or nuclear clusters due to the screening of Coulomb energies of background 
electrons as well as corrections due to excited states and excluded volume 
effects are taken into account in this calculation.  

The total canonical partition function of the in-homogeneous matter is given by,
\begin{eqnarray}
Z(T,V,\{N_i\})=Z_{nuc}~\prod_{A,Z}Z_{A,Z}~Z_{Coul} .
\end{eqnarray}
Here $Z_{nuc}$, $Z_{A,Z}$, $Z_{Coul}$ represent partition functions 
corresponding to the contributions of unbound nucleons, nuclei and Coulomb 
interaction, respectively.

The free energy density is defined as
\begin{eqnarray}
f&=&\sum_{A,Z} f_{A,Z}^0(T,n_{A,Z})+f_{Coul}(n_e,n_{A,Z})+\xi f_{nuc}^0(T,n'_n,n'_p)-T\sum_{A,Z} n_{A,Z} \mathrm{ln}{\kappa}\; ,
\label{fe}
\end{eqnarray}
where the first term gives the contribution of non-interacting nuclei, 
$f_{Coul}$ corresponds to the Coulomb energy, the contribution of interacting
nucleons $f_{nuc}^0$ is multiplied by the available volume fraction of nucleons
$\xi$, $n_n^{'}$ and $n_p^{'}$ are local neutron and proton number densities 
and the last term
goes to infinity when available volume fraction of nuclei ($\kappa$) is
zero near the saturation density. The number density of nuclei is given by the 
modified Saha equation \cite{apj14,hs},
\begin{eqnarray}
&&n_{A,Z}=\kappa~g_{A,Z}(T)\left(\frac{M_{A,Z} T}{2\pi}\right)^{3/2}\exp\left(\frac{(A-Z)\mu_{n}^0+Z\mu_{p}^0-M_{A,Z}-E^{Coul}_{A,Z}-P^0_{nuc}V_{A,Z}}T\right) \; , \label{eq_naz}
\end{eqnarray}
where the meaning of different quantities in the equation can be found from 
Ref.\cite{apj14,hs}. Finally, the pressure is calculated as mentioned in 
Ref.\cite{apj14,hs}. 

\subsection{Density dependent field theory for dense matter}
We calculate the EoS of uniform matter above the saturation density at finite
temperature within the frame of a density dependent relativistic hadron field 
theory \cite{typ,apj14}. In
this case, the dense matter is made of neutrons, protons, hyperons and 
electrons. Being the lightest hyperons, $\Lambda$ hyperons populate the dense 
matter first. Furthermore, heavier hyperons such as $\Sigma$ and $\Xi$ are 
excluded from this calculation because very little is known about their 
interaction in nuclear medium experimentally. The starting point here is the
Lagrangian density for baryon-baryon interaction mediated by the exchange of
$\sigma$, $\omega$ and $\rho$ mesons. The interaction among $\Lambda$ hyperons
is taken into account by the exchange of $\phi$ mesons \cite{apj14} as 
described by the Lagrangian density, 
\label{subsec:2}

\begin{eqnarray}
\label{eq_lag_b}
{\cal L}_B &=& \sum_B \bar\psi_{B}\left(i\gamma_\mu{\partial^\mu} - m_B
+ g_{\sigma B} \sigma - g_{\omega B} \gamma_\mu \omega^\mu 
- g_{\phi B} \gamma_\mu \phi^\mu 
-  g_{\rho B} 
\gamma_\mu{\mbox{\boldmath $\tau$}}_B \cdot 
{\mbox{\boldmath $\rho$}}^\mu  \right)\psi_B\nonumber\\
&& + \frac{1}{2}\left( \partial_\mu \sigma\partial^\mu \sigma
- m_\sigma^2 \sigma^2\right)
-\frac{1}{4} \omega_{\mu\nu}\omega^{\mu\nu}\nonumber\\
&&+\frac{1}{2}m_\omega^2 \omega_\mu \omega^\mu
-\frac{1}{4} \phi_{\mu\nu}\phi^{\mu\nu}
+\frac{1}{2}m_\phi^2 \phi_\mu \phi^\mu \nonumber\\
&&- \frac{1}{4}{\mbox {\boldmath $\rho$}}_{\mu\nu} \cdot
{\mbox {\boldmath $\rho$}}^{\mu\nu}
+ \frac{1}{2}m_\rho^2 {\mbox {\boldmath $\rho$}}_\mu \cdot
{\mbox {\boldmath $\rho$}}^\mu.
\end{eqnarray}
Here $\psi_B$ denotes the baryon octets, ${\mbox{\boldmath 
$\tau_{B}$}}$ is the isospin operator and $g$s are density dependent
meson-baryon couplings. It is to be noted that $\phi$ mesons are mediated
among $\Lambda$ hyperons only.
The pressure is given by \cite{apj14},
\begin{eqnarray}
P &=& -\frac{1}{2}m_\sigma^2 \sigma^2
+ \frac{1}{2} m_\omega^2 \omega_0^2 
+ \frac{1}{2} m_\rho^2 \rho_{03}^2  
+ \frac{1}{2} m_\phi^2 \phi_0^2 
+ \Sigma^r \sum_{B=n,p,\Lambda} n_B
\nonumber \\
&& + 2T \sum_{i=n,p,\Lambda} \int \frac{d^3 k}{(2\pi)^3} 
[ln(1 + e^{-\beta(E^* - \nu_i)}) +
ln(1 + e^{-\beta(E^* + \nu_i)})] ~,  
\end{eqnarray}

where the temperature is defined as $\beta = 1/T$ and
$E^* = \sqrt{(k^2 + m_i^{*2})}$.
This involves 
the rearrangement term $\Sigma^{r}$ \cite{apj14,hof} due to many-body
correlations which is given by
\begin{equation}\label{eq_rear}
\Sigma^{r}=\sum_B[-g_{\sigma B}'  
\sigma n^{s}_B + g_{\omega B}' \omega_0 n_B
+ g_{\rho B}'\tau_{3B} \rho_{03} n_B
+   g_{\phi  B}' \phi_0 n_B  ]~,
\end{equation}
where $'$ denotes derivative with respect to baryon density of species B.

The energy density is
\begin{eqnarray}
\epsilon &=& \frac{1}{2}m_\sigma^2 \sigma^2
+ \frac{1}{2} m_\omega^2 \omega_0^2 
+ \frac{1}{2} m_\rho^2 \rho_{03}^2  
+ \frac{1}{2} m_\phi^2 \phi_0^2 
\nonumber \\
&& + 2 \sum_{i=n,p,\Lambda} \int \frac{d^3 k}{(2\pi)^3} E^* 
\left({\frac{1}{e^{\beta(E^*-\nu_i)} 
+ 1}} + {\frac{1}{e^{\beta(E^*+\nu_i)} + 1}}\right)~.  
\end{eqnarray}

Parameters of the Lagrangian density are computed using available experimental 
data at the saturation density. Meson-nucleon couplings are determined by 
fitting the properties of finite nuclei using some functional forms of density
dependent couplings \cite{typ}. This parameter set is known as the DD2. For 
vector meson couplings of $\Lambda$ hyperons, we
exploit the SU(6) symmetry relations whereas the scalar coupling is obtained 
from $\Lambda$ hypernuclei data with a potential depth of $-30$ MeV at the
saturation density \cite{sch96}.

The descriptions of non-uniform and uniform matter are matched at the 
crust-core boundary in a thermodynamically consistent manner \cite{apj14}.
Charge neutrality and $\beta$-equilibrium conditions are imposed for neutron
star matter.   

\begin{figure}[th]
\vspace{1cm}
\centerline{\includegraphics[width=6.0cm]{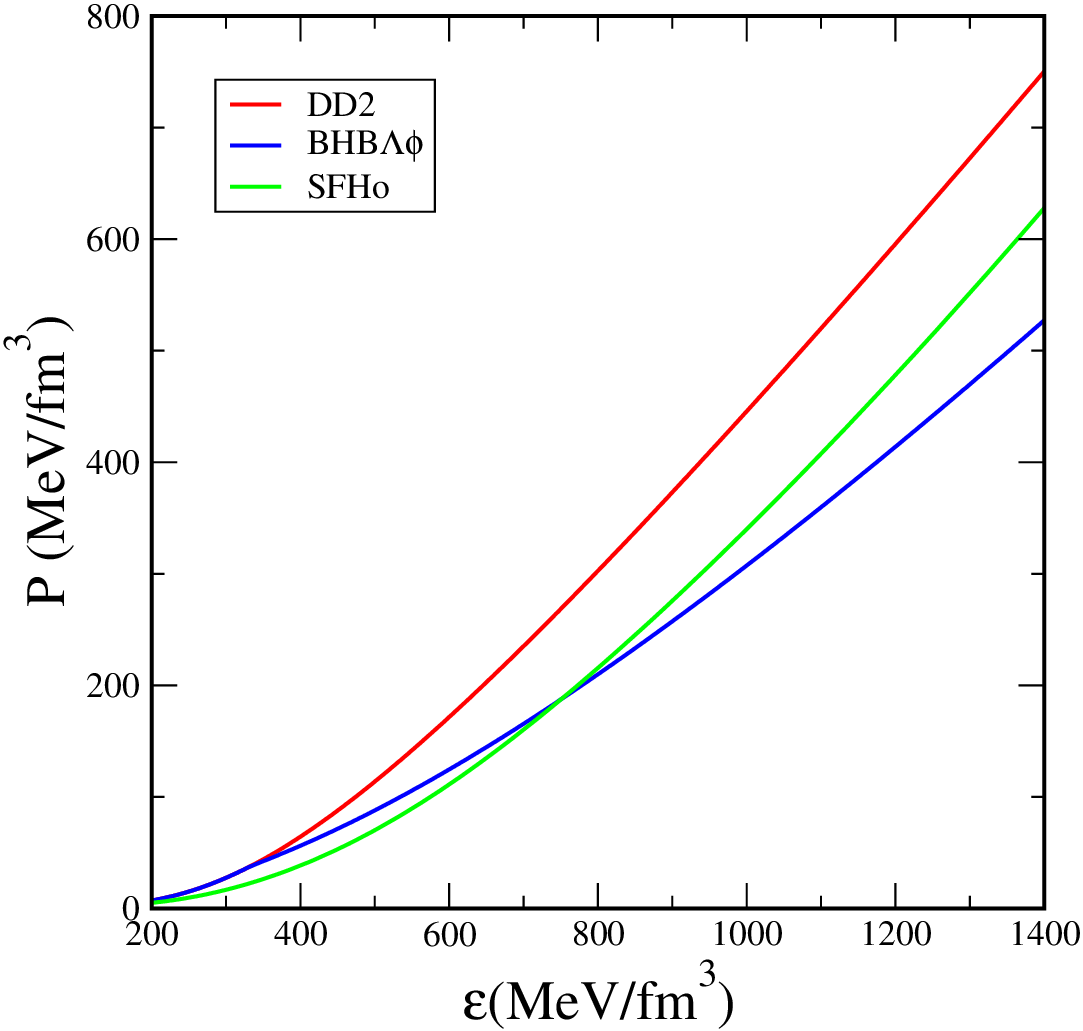}}
\caption{Pressure versus energy density (EoS) is shown for the DD2, 
BHB$\Lambda\phi$ and SFHo EoS models.}
\label{fig1}       
\end{figure}
\begin{figure}[th]
\centerline{\includegraphics[width=5cm]{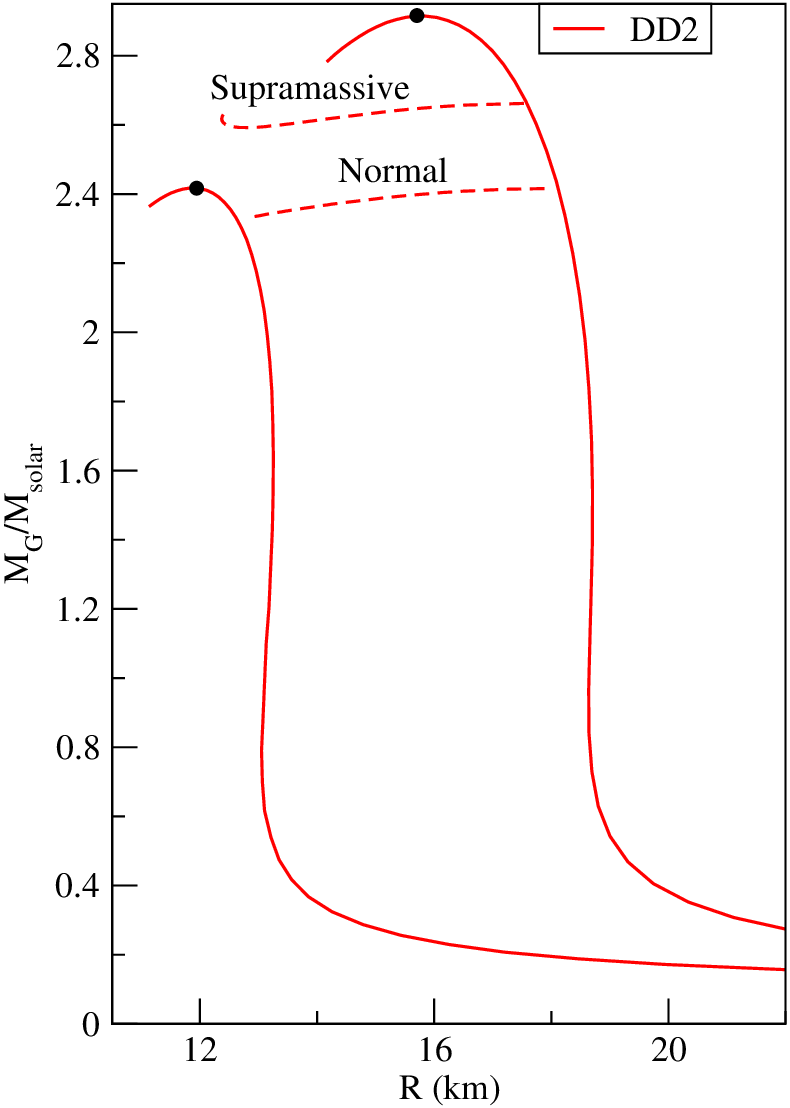}\hspace{0.5cm}\includegraphics[width=5cm]{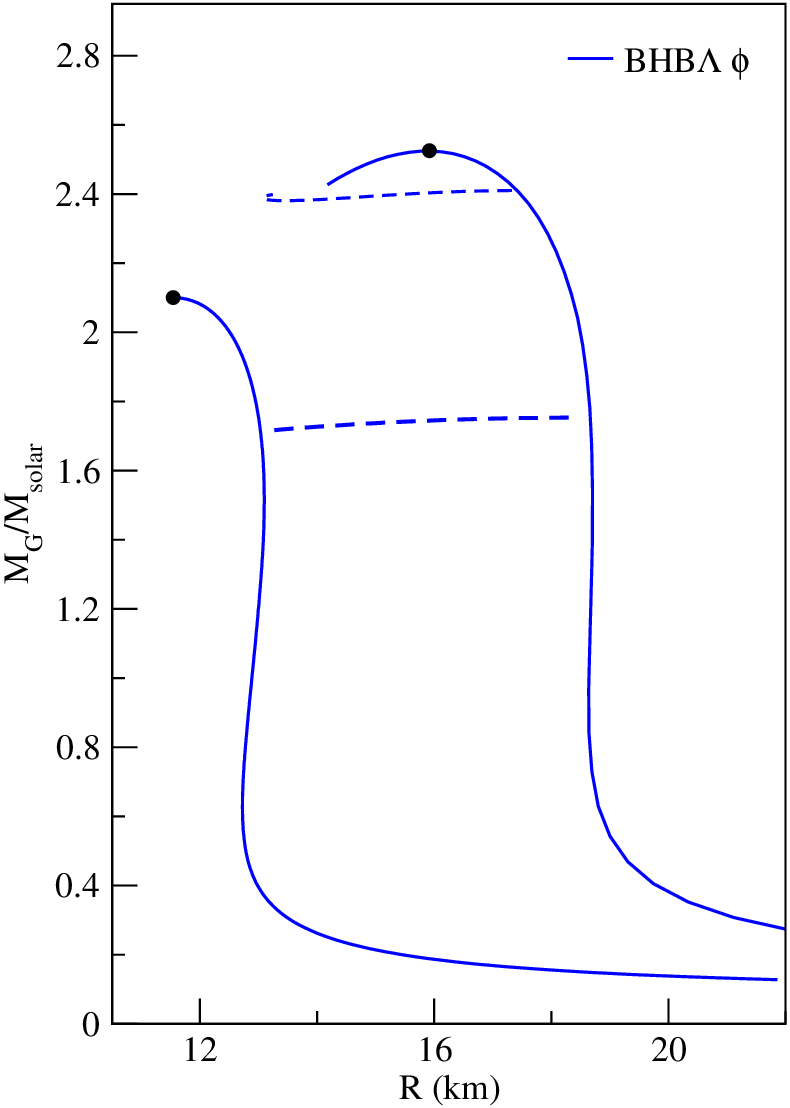}}
\caption{Mass-radius relationship is shown for the DD2 EoS in left panel and
BHB$\Lambda\phi$ in the right panel. In both panels, the bottom curve 
represents the non-rotating sequence and the upper curve corresponds to the
sequence of neutron stars uniformly rotating at their Keplerian frequencies.}
\label{fig2}       
\end{figure}

\section{Maximum Mass of Neutron Star}
\label{sec:3}
Here we discuss the results of our calculation. As discussed in the preceding
section, we consider neutron star matter made of neutrons, protons, $\Lambda$ 
hyperons and electrons in the DDRH model. The EoS corresponding to nucleons 
only matter is denoted as the DD2 whereas the EoS of dense matter involving 
$\Lambda$ 
hyperons is known as the BHB$\Lambda\phi$. We also include the SFHo nuclear EoS
of Steiner et al. in this discussion \cite{stein}. Figure 1 displays the EoSs
(pressure versus energy density) corresponding to the DD2, BHBH$\Lambda\phi$ 
and SFHo models. It shows that the DD2 EoS is the stiffest among the three. 
Further we 
note that the SFHo EoS was softer over a certain region of energy density but 
becomes stiffer at higher densities than the BHB$\Lambda\phi$. However, it
follows from the structure calculation using the Tolman-Oppenheimer-Volkoff 
(TOV) equation that the overall SFHo EoS is softer 
compared with the BHB$\Lambda\phi$ EoS. Maximum masses of non-rotating neutron
stars are 2.42, 2.11 and 2.06 M$_{\odot}$ corresponding to the DD2, 
BHB$\Lambda\phi$ and SFHo EoS, respectively. All these EoSs are compatible with
the observed 2 M$_{\odot}$ neutron star \cite{anto}.
    
We also compute the structures of rotating neutron stars using the LORENE 
library \cite{eric,lorene}. Mass-radius relationships of (non)-rotating neutron 
stars are exhibited in Fig. 2. The sequences of non-rotating
neutron stars (bottom curve) and uniformly rotating neutron stars (upper curve)
at Keplerian frequencies are plotted for the DD2 EoS in the left panel and for
the hyperon EoS BHB$\Lambda\phi$ in the right panel. Horizontal lines in both
panels are fixed rest mass sequences. Those are denoted as normal and 
supramassive sequences. Rotating neutron stars evolve along 
those sequences keeping the total baryon mass conserved. The normal sequence
finds its counterpart on the non-rotating star branch spinning down whereas 
neutron stars following the supramassive sequence would finally collapse into 
black holes. Any evolutionary sequence above the maximum mass rotating 
neutron star is known as the hypermassive sequence and a neutron star in this
sequence would be stabilised only by differential rotation before collapsing 
into a
black hole in  a few tens of milli seconds. Recently, it was demonstrated that 
the relation between the maximum mass ($M_{max}^{Rot}$) of the rotating neutron 
star at the Keplerian frequency and that ($M_{max}^{TOV}$) of the non-rotating 
neutron star satisfied a universal relation \cite{breu,ssl}. This relation is
given by \cite{breu} 
\begin{equation}
M_{max}^{Rot} = 1.203 \pm 0.022 M_{max}^{TOV}~.
\end{equation}
      
With this understanding of different evolutionary sequences respecting total
baryon mass conservation, we discuss the fate of the massive remnant 
formed in merger event GW170817. The remnant could not be a hypermassive 
neutron star undergoing a prompt collapse to a black hole because a large
amount of ejected matter was observed in the event \cite{metz}. This implies 
that the massive remnant existed for some duration. However, a long 
lived massive remnant is ruled out because of a sGRB sighted 1.7 s after the 
merger.
It is inferred that the massive remnant collapsed to a black hole close
to the maximum mass of a uniformly rotating sequence \cite{rezo,shap}. This
description might be intimately tied to the maximum mass of the non-rotating 
neutron star. 

It is estimated from the observation of neutron star merger event GW170817 
assuming low dimensionless spins for the neutron stars in the binary that the 
total binary mass was $\sim$ 2.74 M$_{\odot}$. The mass loss 
from the merged
object due to emissions of gravitational waves and neutrinos and ejected 
neutron rich matter amounts to $\sim$ 0.15 $\pm 0.03$ M$_{\odot}$ 
\cite{shib}. Consequently, the mass of the remnant reduced 
to $\sim$ 2.6 M$_{\odot}$. If we identify this mass of the remnant that might 
have collapsed into a black hole, with the maximum mass of the uniformly 
rotating neutron star at the Keplerian frequency i.e.
M$_{max}^{Rot}$ of Eq. (8), an upper limit on the maximum mass of non-rotating
neutron stars might be obtained \cite{rezo}. It follows from Eq. (8) that the 
upper limit 
is $\sim$ 2.16 M$_{\odot}$. It is already known from the observations of 
galactic pulsars that the lower limit on the maximum mass of non-rotating 
neutron 
stars is 2.01 M$_{\odot}$. All these information put together lead to 
\begin{eqnarray}
2.01 M_{\odot} \leq M_{max}^{TOV} < 2.16 M_{\odot}~.
\end{eqnarray}
Different groups converged almost to the same value of the upper limit from 
different analyses of GW170817 \cite{metz,rezo,shap,shib}. 
 
\begin{figure}[th]
\vspace{0.7cm}
\centerline{\includegraphics[width=5.5cm]{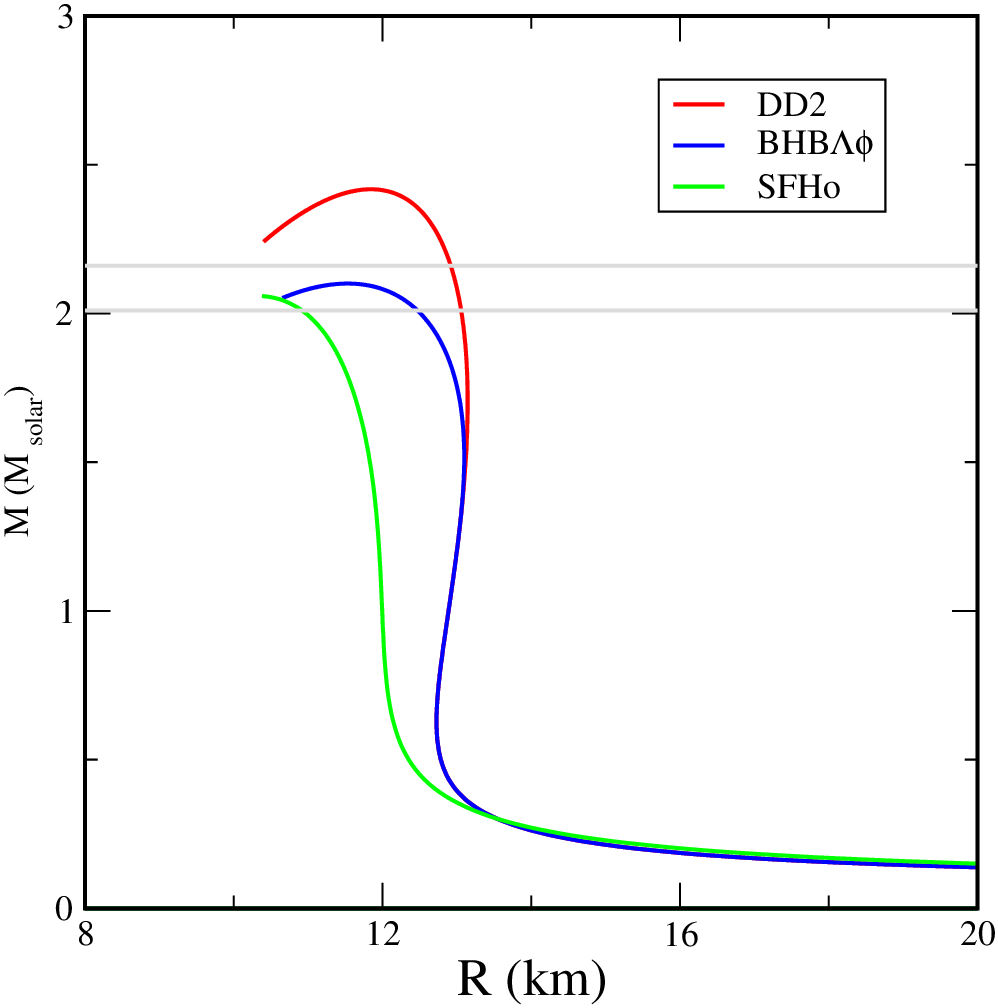}}
\caption{Mass-radius relationships of non-rotating neutron stars are shown for
the DD2, BHB$\Lambda\phi$ and SFHo EoSs. Horizontal lines denote the lower 
bound and upper bound on the maximum mass as given by Eq. (9).}
\label{fig3}       
\end{figure}

\subsection{Constraint on EoS}
We discuss the implications of the lower and upper limits of the maximum mass
on EoSs. 
Mass-radius relationships corresponding to the DD2, BHB$\Lambda\phi$ and SFHo 
EoSs are plotted in Fig. 3. The lower and upper limits on the maximum mass are 
also indicated by two horizontal lines. It is evident from the figure that the 
BHB$\Lambda\phi$ and SFHo EoSs are consistent with both limits of the maximum 
mass.
But this is not case with the DD2 EoS because it fails to satisfy the upper 
limit. It is to be noted that the DD2, BHB$\Lambda\phi$ and SFHo EoSs are 
being used for neutron star merger simulations by various groups 
\cite{rad17,shib}.    

Next we perform a comparative study of different EoSs. Particularly, we look at
the nuclear matter properties of all EoSs such as the saturation density 
($n_0$), binding energy(E$_{0}$), incompressibility (K), symmetry energy (S) 
and its density slope (L). The nuclear matter properties of eight 
EoSs are recorded in Table 1. The last row of the table gives experimental 
values of nuclear matter properties \cite{hem}. 
First five of those EoSs for example, Lattimer-Swesty 200 (LS200) \cite{ls}, 
Skyrme Lyon (SLy) \cite{han}, M\"uller-Serot 1 (MS1) \cite{ms}, 
Akmal-Pandharipande-Ravenhall 4 (APR4) \cite{apr} and hyperon EoS H4 \cite{owe}
were used in the analysis of GW170817 \cite{abbott} because all of them satisfy
the lower limit on the maximum mass. It is to be noted that all are nucleons
only EoSs except the H4 EoS. A closer look at the nuclear matter 
properties at the saturation density of first five EoSs throw up important
information about their behaviour at higher densities. It is evident from the
Table that one or more observables of nuclear matter in case of LS220, MS1, 
APR4 and H4 EoSs are not consistent with the experimental values. This leads 
to a very
soft or stiff EoS in those cases. For example, high values of incompressibility
(K) for the APR4 and H4 make them stiffer EoSs. The threshold for the 
appearance of hyperons is shifted to a lower density for a very stiff EoS 
leading to the large population of hyperons in dense matter and resulting in a 
lower maximum mass neutron star as it is happening in case of the H4 EoS. For 
the LS220 and MS1 EoSs, the density slope (L) of 
symmetry energy is much higher than the experimental value. As a result, the 
maximum mass for the MS1 
EoS is higher than the upper limit of 2.16 M$_{\odot}$. However, the 
interplay between a lower value of K and higher value of L for the LS220 EoS 
determines the maximum mass which falls well within the limits of Eq. (9). 
Though the SLy EoS is consistent with the experimental values and observational 
limits on the maximum mass, it is a non-relativistic EoS and superluminal 
behaviour could be a problem in this case at very high density (5-8 $n_0$) 
\cite{apr}. We have already discussed the last three EoSs of the Table. The 
nuclear matter 
properties of the DD2, SFHo and BHB$\Lambda\phi$ EoSs are in good agreement 
with the experimental values. However, it is concluded that the DD2 EoS is 
ruled out by Eq. (9). It is possible to further constrain EoSs using the 
measured tidal deformability from GW170817. Based on the tidal 
deformability of GW170817, the H4, APR4 and LS220 EoSs are excluded 
whereas the BHB$\Lambda\phi$ EoS is consistent with GW170817 data \cite{rad2}. 

\begin{table}
\caption{Nuclear matter properties of different EoSs used in the analysis of 
GW170817 \cite{abbott} and in this article are recorded here. Experimental 
values of saturation density ($n_0$), binding energy ($E_{0}$), 
incompressibility ($K$), symmetry energy ($S$) and its density slope ($L$) are 
listed in the last row of this table \cite{hem}. Maximum mass of non-rotating 
neutron stars 
corresponding to each EoS is also shown here. Lower bound on the maximum mass
of non-rotating neutron stars is mentioned in the last row \cite{anto}.}
\label{tab:1}       
%
\begin{tabular}{p{2cm}p{1.5cm}p{1.5cm}p{1.5cm}p{1.5cm}p{1.5cm}p{1.5cm}}
\hline\noalign{\smallskip}
EoS&$n_0$& E$_{0}$& K & S & L & M$_{max}$\\
    & {[fm$^{-3}$]} & [MeV] & [MeV] &[MeV] & [MeV]
& [M$_{\odot}$] \\
\noalign{\smallskip}\svhline\noalign{\smallskip}
LS220 & 0.1550  &16.00 & 220 & 28.61 & 73.82 & 2.06\\
SLy & 0.160  &15.97 & 230 & 32.00 & 45.94 & 2.05\\
MS1 & 0.1484 &15.75 & 250 & 35.00 & 110.00 & 2.77\\
APR4 & 0.160 &16.00 & 266 & 32.59 & 58.46 & 2.19\\
H4 & 0.153  & 16.3  & 300 &  32.5 & 94.02 & 2.02\\
DD2& 0.1491 & 16.02 & 243 & 31.67 & 55.04 & 2.42\\
SFHo & 0.1583 & 16.19 & 245 & 31.57 & 47.10 & 2.06\\
BHB$\Lambda\phi$& 0.1491 & 16.02 & 243 & 31.67 & 55.04 & 2.11\\
Exp. & $\sim 0.15$ & $\sim 16$ & $240\pm10$ & $29.0-32.7$ & $40.5-61.9$ &2.01$\pm$0.04\\
\noalign{\smallskip}\hline\noalign{\smallskip}
\end{tabular}
\end{table}


\section{Summary and conclusion}
We have investigated the equations of state of dense matter within the 
framework of the density dependent relativistic hadron field theory. The 
nucleons only EoS is denoted as the DD2 whereas the $\Lambda$ hyperon EoS
is known as the BHB$\Lambda\phi$. The neutron star merger event GW170817
gives an upper limit on the maximum mass of non-rotating neutron stars whereas 
the lower limit is already known from the observations of pulsars. The upper 
and lower limits severely constraint the EoS as we have found through a 
comparative study of eight EoSs and their nuclear matter properties. It is 
found that the BHB$\Lambda\phi$ EoS is consistent with both limits of the 
maximum mass and the tidal deformability of GW170817.

\begin{acknowledgement}

S.B. and D.B. gratefully remember the support and encouragement that they
received always from Professor Walter Greiner.
\end{acknowledgement}

\end{document}